\documentclass[aps,pra,notitlepage,twocolumn,10pt,a4paper]{revtex4-1}

\usepackage{xcolor,graphicx,ulem}
\usepackage{amsmath,amssymb}
\usepackage[colorlinks=true,urlcolor=blue,citecolor=blue,linkcolor=blue]{hyperref}

\usepackage{xr}

\begin{document}

\title{Quantum non-Gaussianity and quantification of nonclassicality}
\author{B. K\"uhn}\email{benjamin.kuehn2@uni-rostock.de}
\author{W. Vogel}
\affiliation{Arbeitsgruppe Quantenoptik, Institut f\"ur Physik, Universit\"at Rostock, D-18051 Rostock, Germany}

\date{\today}

\begin{abstract}
	The algebraic quantification of nonclassicality, which naturally arises from the quantum superposition principle, is related to properties of regular nonclassicality quasiprobabilities. 
	The latter are obtained by non-Gaussian filtering of the Glauber-Sudarshan $P$~function. 
	They yield lower bounds for the degree of nonclassicality. 
	We also derive bounds for convex combinations of Gaussian states for certifying quantum non-Gaussianity directly from the experimentally accessible nonclassicality quasiprobabilities. 
	Other quantum-state representations, such as $s$-parametrized quasiprobabilities, insufficiently indicate or even fail to directly uncover detailed information on the properties of quantum states.	
	As an example, our approach is applied to multi-photon-added squeezed vacuum states.
\end{abstract}

\maketitle

\section{Introduction}
\label{ch:introduction}
	Uncovering structural information on quantum states is of fundamental importance for present quantum technologies.
	A prominent example for such characteristics is the classification of a state as classical or nonclassical \cite{Titulaer1965}.
	Beyond this bivalent categorization, the amount and the kind of nonclassicality provided by various quantum resources is of great interest to figure out optimal experimental implementations of quantum technologies.
	In particular, a further state property---the quantum non-Gaussianity \cite{Barbieri2010,Jezek2011,Filip2011,Genoni2013}---is important for various applications in quantum technology \cite{Straka2014,Lasota2017,Straka2018}.
			
	In the past decades distinct measures for the quantification of nonclassicality have been proposed for harmonic oscillator quantum systems, such as light, the quantized motion of trapped atoms, and others.
        The early attempts rely on topological properties of quantum states; they are based on the distance of a state to the set of classical states. 
        Such an approach can be based on different notions of the distance, such as the trace-norm-induced distance \cite{Hillery1987}, the Monge distance \cite{Monge1998}, or the Hilbert-Schmidt distance \cite{Dodonov2009}. 
        The crucial point of such quantifications consists in its ambiguity. 
        On the other hand, a quantification based on the number of superpositions of coherent states relies on the fundamental quantum superposition principle and yields an unambiguous quantification of nonclassicality~\cite{SpVo2015}.
       	Note that distance-based measures have also been studied for the quantification of non-Gaussianity \cite{Genoni2007,Genoni2008,Park2017}.
	
	Alternatively, the quantification problem has been addressed by using properties of quasiprobabilities~\cite{Lee1991,Luetkenhaus1995}, which are full representations of the quantum states \cite{Cahill1969,Agarwal1970}.
	They may certify nonclassicality of various states~\cite{Wineland1996,Haroche2008}. 
	A powerful tool is the nonclassicality quasiprobability~\cite{Kiesel2010}---a regularized form of the Glauber-Sudarshan $P$~function~\cite{Sudarshan1963,Glauber1963}. 
	It is designed such that it identifies any nonclassical state through negativities. 
	This could be experimentally shown even for the strongly singular squeezed states~\cite{Kiesel2011_2,Agudelo2015}.
	In Ref. \cite{Asboth2005}, the potential of a single-mode nonclassical state to generate quantum entanglement \cite{Horodecki2009} by applying only linear optical elements was used to quantify nonclassicality.
	An algebraic approach for the quantification of nonclassicality was introduced, defining the degree of nonclassicality by the number of quantum superpositions of coherent states~\cite{Gehrke2012}. 
	Based on the Schmidt number, an algebraic quantification was also used for quantum entanglement~\cite{Terhal2000,Sanpera2001,Horodecki2009}, which can be generalized to multipartite scenarios~\cite{Eisert2001}. 
	A unified quantification of nonclassicality and entanglement has been introduced~\cite{Sperling2014}, which directly relates the degree of nonclassicality to the Schmidt number of bipartite and even multipartite entangled states created by linear optical devices, such as an $N$ splitter.  
 	This is an important relation as entanglement is the basis for quantum technologies such as quantum information processing \cite{Nielsen2000} and secure communication protocols \cite{Gisin2007}.
 	For practical applications, the degree of nonclassicality can be verified both by a witness~\cite{Mraz2014} and through the structure of the characteristic function of the Glauber-Sudarshan $P$~function~\cite{Ryl2017}. 
	However, these methods have not been developed for the purpose of uncovering quantum non-Gaussianity.
				
	In the present paper we prove the usefulness of nonclassicality quasiprobabilities for certifying both quantum non-Gaussianity and the degree of nonclassicality.
	The single class of phase-space distributions under study uncovers both topological and algebraic quantifications of nonclassicality.
	We may uncover dissimilar features of arbitrary quantum states, such as nonclassicality, non-Gaussianity, and the minimal number of quantum superpositions forming a given state. 
	As the required phase-space functions are experimentally accessible, the present method also applies to state-of-the-art experiments.
		
\section{Nonclassicality versus Quantum Non-Gaussianity}
\label{ch:quantification}

	The coherent states $|\alpha\rangle$ are well known to resemble the classical behavior of the harmonic oscillator and, hence, the classical character of light.
	Since mixing is a classical operation, a general state $\hat\rho$ is referred to as classical if it can be written as a mixture of coherent states,
	\begin{align}
		\hat\rho=\int d^2\alpha\,P_{\mathrm{cl}}(\alpha)\,|\alpha\rangle\langle\alpha|,
	\end{align}
	with a classical probability distribution $P_{\mathrm{cl}}$.
	In other words, classical states are elements of the convex hull of the set of coherent states. 
	In fact, any quantum state can be written in the coherent state basis \cite{Glauber1963,Sudarshan1963},
	\begin{align}
		\hat\rho=\int d^2\alpha\,P(\alpha)\,|\alpha\rangle\langle\alpha|.
	\end{align}
	However, the Glauber-Sudarshan function $P(\alpha)$ may fail to show the properties of a classical probability.
	Such states are referred to as nonclassical ones~\cite{Titulaer1965}.
	On this basis, any nonclassicality requires quantum superpositions of coherent states,
	\begin{align}\label{eq:psir}
		|\psi_r\rangle=\sum_{j=1}^r\mu_j|\gamma_j\rangle,
	\end{align}
	where $\mu_j$ are nonzero complex numbers, $|\gamma_j\rangle$ are various coherent states, and $r$ is the number of superpositions.
        Now we may define the degree of nonclassicality as 
        \begin{align}\label{eq:ncldegree}
		\kappa=r-1
	\end{align}	
	or as a monotonous function of this quantity~\cite{Gehrke2012}.
	This is an ambiguous quantification of nonclassicality, as an increase of nonclassicality is caused by a larger number of quantum superpositions.
	Pure states with nonclassicality degree $r-1$ can be written in the form \eqref{eq:psir} with the minimal number of $r$ superpositions of coherent states.
	
	Extending these considerations, a general quantum state $\hat\rho$ has a nonclassicality degree $r-1$ if it is a classical mixture of pure states of a nonclassicality degree of at most $r-1$, i.e.,
	\begin{align}
		\hat\rho=\int dP_{\mathrm{cl}}(|\psi_r\rangle)|\psi_r\rangle\langle\psi_r|.
	\end{align}
	In the case $r>1$, different such decompositions exist; thus, the nonclassicality degree refers to the minimal possible $r$. 
	The states with a degree of nonclassicality of at most $r-1$ form a closed, convex set $\mathcal{M}_r$.
	This definition is straightforwardly generalized to multimode harmonic oscillator systems; see also Ref. \cite{Mraz2014}.
	
	A further important characteristics of quantum states is the non-Gaussianity.
	The Gaussian state $|G_{\boldsymbol{u},\boldsymbol{\Sigma}}\rangle$ is fully determined through the first and second moment, namely, the mean value $\boldsymbol{u}$ and the covariance matrix $\boldsymbol{\Sigma}$.
	In principle, there are two types of non-Gaussianity. 
	On the one hand, a non-Gaussian state can be obtained by properly mixing Gaussian states according to a classical probability distribution $P_{\mathrm{cl}}$, i.e.,
	\begin{align}\label{eq:GaussHull}
		\hat\rho=\int dP_{\mathrm{cl}}(\boldsymbol{u},\boldsymbol{\Sigma})\,|G_{\boldsymbol{u},\boldsymbol{\Sigma}}\rangle\langle G_{\boldsymbol{u},\boldsymbol{\Sigma}}|.
	\end{align}
	These states form also a closed, convex set $\mathcal{G}$ (see Ref.~\cite{Genoni2013}).
	However, for quantum technologies this kind of non-Gaussianity is rather useless, since it originates from classical noise, such as phase randomization of Gaussian quantum states \cite{Franzen2006}. 
	Of greater interest are quantum non-Gaussian states, $\hat\rho\not\in\mathcal{G}$, whose non-Gaussianity is intrinsically quantum.

\section{Structural information in phase-space functions}
\label{ch:unification}

	It is possible to use a witness approach to formulate criteria for the quantification of nonclassicality.
	A corresponding method was proposed \cite{Mraz2014}, where the lower and upper bounds, $\overline{g}_r$ and $\underline{g}_r$, of the expectation value of a given Hermitian operator $\hat L$ are determined with respect to $\mathcal{M}_r$.
	If $\langle\hat L\rangle>\overline{g}_r$ or $\langle\hat L\rangle<\underline{g}_r$ for an unknown state, the nonclassicality degree of this state is shown to be necessarily greater than or equal to $r$.
	Equivalently, one introduces witness operators $\hat {\overline{W}}_r=\overline{g}_r\hat 1-\hat L$ and $\hat{\underline W_r}=\hat L-\underline{g}_r\hat 1$, such that $\langle\hat {\overline{W}}_r\rangle\geq 0$ and $\langle\hat {\underline{W}}_r\rangle\geq 0$ for all states in $\mathcal{M}_r$. 
	
	For the case $r=1$, a complete family of witness operators, $\{\hat W\}$, is already known \cite{Kiesel2012}, which is able to show that any state with nonclassicality degree $r-1>0$ (nonclassical state) cannot be written as a convex combination of pure nonclassicality degree-zero states (coherent states).
	Thus, these witnesses uncover all nonclassical effects of single- and multimode harmonic oscillator systems.
	In the single-mode case, they are of the form 
	\begin{align}\label{eq:Wwalpha}
		\hat W_{w,\alpha}=\hat D^\dagger_\alpha\hat W_w\hat D_\alpha
	\end{align}
	with 	
	\begin{align}\label{eq:Ww}
		\hat W_w=\dfrac1{\pi^2}\int d^2\beta\,\Omega_w(\beta)\,e^{\beta\hat a^\dagger}e^{-\beta^\ast\hat a},
	\end{align}
	and the coherent displacement operator $\hat D_\alpha=\exp\left[\alpha\hat a^\dagger-\alpha^\ast\hat a\right]$.
	Here $\hat a$ and $\hat a^\dagger$ are the bosonic annihilation and creation operators, respectively.
	In total, there are only three free real parameters, a positive quantity $w$ and a complex number $\alpha$.
	The so-called nonclassicality filter $\Omega_w$ is chosen in such a way that the expectation value of $\hat W_{w,\alpha}$ exists for all states.
	Furthermore, its specific structure guarantees that this expectation value is non-negative for all classical states.
	A necessary and sufficient condition for a state being nonclassical is the existence of parameters $w$ and $\alpha$ such that $\langle\hat W_{w,\alpha}\rangle<0$.
	
	The witnesses $\hat W_{w,\alpha}$ are associated with phase-space functions, the latter being an established method to visualize quantum effects through their negativities.
	In particular, the expectation value $P_w(\alpha)=\langle\hat W_{w,\alpha}\rangle$ can be regarded as a nonclassicality quasiprobability in phase space; it holds $\int d^2\alpha\,P_w(\alpha)=1$. 
	For any positive value $w$ this function is nonsingular, as $\hat W_{w,\alpha}$ is a bounded operator, and, thus, it is in principle accessible in experiments.
	In the limit $w\to\infty$ the function $P_w$ approaches the Glauber-Sudarshan $P$ function, which is our reference for nonclassicality.
	Note that $s$-parametrized quasiprobabilities \cite{Cahill1969}, which correspond to a Gaussian filter $\Omega_w$ in Eq. \eqref{eq:Ww}, are only regular if the $s$ parameter is sufficiently small for the state under study. 
	There is no $s$-parametrized quasiprobability which visualizes the nonclassicality of a squeezed vacuum state.
	Introducing non-Gaussian filters that decay more strongly than any Gaussian and that have a non-negative Fourier transform resolve this problem.
	
	\begin{table}[t]
		\centering
		\caption{For parameters $w$ in the range $[w_{\mathrm{min}},w_{\mathrm{max}}]$ in each column, a lower bound for the degree of nonclassicality of up to $\kappa$ can be verified.
		In this range the overall supremum (infimum) of the nonclassicality quasiprobability $P_w(\alpha)$ is attained by a Fock state $|\overline{n}\rangle$ ($|\underline{n}\rangle$).
		}
		\label{tab:wc}
		\begin{tabular}{c|cccccccc}
		\hline\hline
		$w_{\mathrm{min}}$&1.200&1.550&1.795&2.027&2.239&2.436&$\cdots$\\
		$w_{\mathrm{max}}$&1.550&1.795&2.027&2.239&2.436&2.619&$\cdots$\\
		\hline
		$\overline{n}$&0&2&2&4&4&6&$\cdots$\\
		$\underline{n}$&1&1&3&3&5&5&$\cdots$\\
		$\kappa$&1&2&3&4&5&6&$\cdots$\\
		\hline\hline
		\end{tabular}
	\end{table}
	
	The witnesses $\hat W_{w,\alpha}$ in Eq. \eqref{eq:Wwalpha} together with Eq. \eqref{eq:Ww} and, therefore, the associated nonclassicality quasiprobabilities provide a full test for nonclassicality.
	However, it is unknown yet to what extent structural state characteristics, such as the nonclassicality degree and quantum non-Gaussianity, can also be uncovered solely on the basis of these specific quantities.
	In order to approach this question, we combine the witness approach for quantifying nonclassicality on the basis of the quantum superposition principle \cite{Mraz2014} and the universal nonclassicality witness operators $\hat W_{w,\alpha}$, corresponding to regular phase-space functions.
	In the following we use the compact support filter defined in Refs. \cite{Kiesel2012,Kuehn2014}, since the associated witness \eqref{eq:Ww} has the closed form expression
	\begin{align}\label{eq:WwFilter}
		\hat W_w&=\,:\dfrac1{\pi}\dfrac{\left[J_1\left(2w\sqrt{\hat a^\dagger\hat a}\right)\right]^2}{\hat a^\dagger\hat a}:\,,
	\end{align}
	where $J_1(\cdot)$ is the Bessel function of the first kind and $:\cdot:$ denotes normal ordering.
	This operator is diagonal in the Fock basis,
 	\begin{align}\label{eq:diagFock}
		\hat W_w=\sum_{n=0}^\infty c_{w,n}|n\rangle\langle n|,
 	\end{align}
 	with the analytical coefficients \cite{Kiesel2012_8}
 	\begin{align}\label{eq:coeff}
 		c_{w,n}=\dfrac{w^2}{\pi}\sum_{m=0}^n\dfrac{(-w^2)^m}{[(m+1)!]^2}\dfrac{n!}{(n-m)!}\binom{2m+2}{m}.
 	\end{align}	
	
	Due to unitarity, the coherently displaced operator $\hat W_{w,\alpha}$ in Eq. \eqref{eq:Wwalpha} has the same eigenvalues as the operator $\hat W_w$.
	Therefore, the overall supremum and infimum of the nonclassicality quasiprobability $P_w(\alpha)$ for a fixed value of $w$ is attained for Fock states $|\overline{n}(w)\rangle$ and $|\underline{n}(w)\rangle$, respectively (see Appendix \ref{ch:totalmaxmin}).
	Table \ref{tab:wc} shows that the supremum is attained for even numbers $\overline{n}$, while the infimum is attained for odd numbers $\underline{n}$.
	These numbers increase sequently and change alternately whenever the parameter $w$ exceeds a critical value.
	Fock states $|n\rangle$ have the nonclassicality degree $\kappa=n$, as they are representable as a quantum superposition of $n+1$ coherent states \cite{Janszky1993},
	\begin{align}\label{eq:fockcoh}
		|n\rangle=\lim_{\epsilon\to 0}\mathcal{C}_n(\epsilon)\sum_{k=0}^n e^{-2\pi ikn/(n+1)}|\epsilon e^{2\pi ik/(n+1)}\rangle,
	\end{align}
	weighted equally and their amplitudes located on a circle in the complex plane with radius tending to zero; $\mathcal{C}_n$ provides the correct normalization.
	This is a maximally nonclassical state in the topological picture.
	Consequently, for given $w$ a nonclassicality degree up to $\max\left[\overline{n}(w),\underline{n}(w)\right]$ can be certified by means of the nonclassicality quasiprobability $P_w(\alpha)$.
	
	Let us compare this result with the case of $s$-parametrized quasiprobabilities.
	In the parameter range $s\leq 0$, where these functions are always regular, the overall supremum is attained by the vacuum state and the overall infimum by the single-photon state.
	In fact, these quasiprobabilities uncover only the presence of some nonclassicality for $s>-1$ without resolving a nonclassicality degree.
	This clearly demonstrates, that the $s$-parametrized quasiprobabilities not only provide an incomplete nonclassicality test; they also do not yield more specific structural insight.
	
	Now, we determine the upper and lower bounds, $\overline{g}_r$ and $\underline{g}_r$, of the nonclassicality quasiprobability $P_w(\alpha)$ associated with the witness operator \eqref{eq:WwFilter} over the set $\mathcal{M}_r$.
	Mixing cannot shift the optimum, thus, the optimization can run over pure states $|\psi_r\rangle$ defined in Eq. \eqref{eq:psir}.
	Furthermore, the coherent displacement in Eq. \eqref{eq:Wwalpha}, can be included in the states.
	In total, the quantity to be optimized reads as
	\begin{align}\label{eq:exp}
		\langle\psi_r|\hat W_{w}|\psi_r\rangle=\dfrac1{\pi}\sum_{\ell=1}^r\sum_{j=1}^r\mu_{\ell}^*\mu_j\left[J_1\left(2w\sqrt{\gamma_\ell^\ast\gamma_j}\right)\right]^2\dfrac{\langle\gamma_\ell|\gamma_j\rangle}{\gamma_\ell^\ast\gamma_j}.
	\end{align}
	The complex coherent amplitudes $(\gamma_1,\dots,\gamma_r)^T$ and complex coefficients $(\mu_1,\dots,\mu_r)^T$ are the optimization parameters, which have to fulfill the normalization constraint $\langle\psi_r|\psi_r\rangle=1$.
	We perform our analysis up to $r=6$, requiring to optimize in total up to $22$ free real parameters, which is done by a genetic search \cite{Haupt2004} together with a gradient descent \cite{Snyman2005}.
	In a similar way, the upper and lower bounds, $\overline{g}_{\mathrm{G}}$ and $\underline{g}_{\mathrm{G}}$, of $P_w(\alpha)$ with respect to the Gaussian convex hull $\mathcal{G}$ are obtained by an optimization over pure Gaussian states.
	The latter are squeezed coherent states, defined by the complex squeezing parameter and complex coherent displacement amplitude.
				
\begin{figure}[t]
		\includegraphics[clip,scale=0.65]{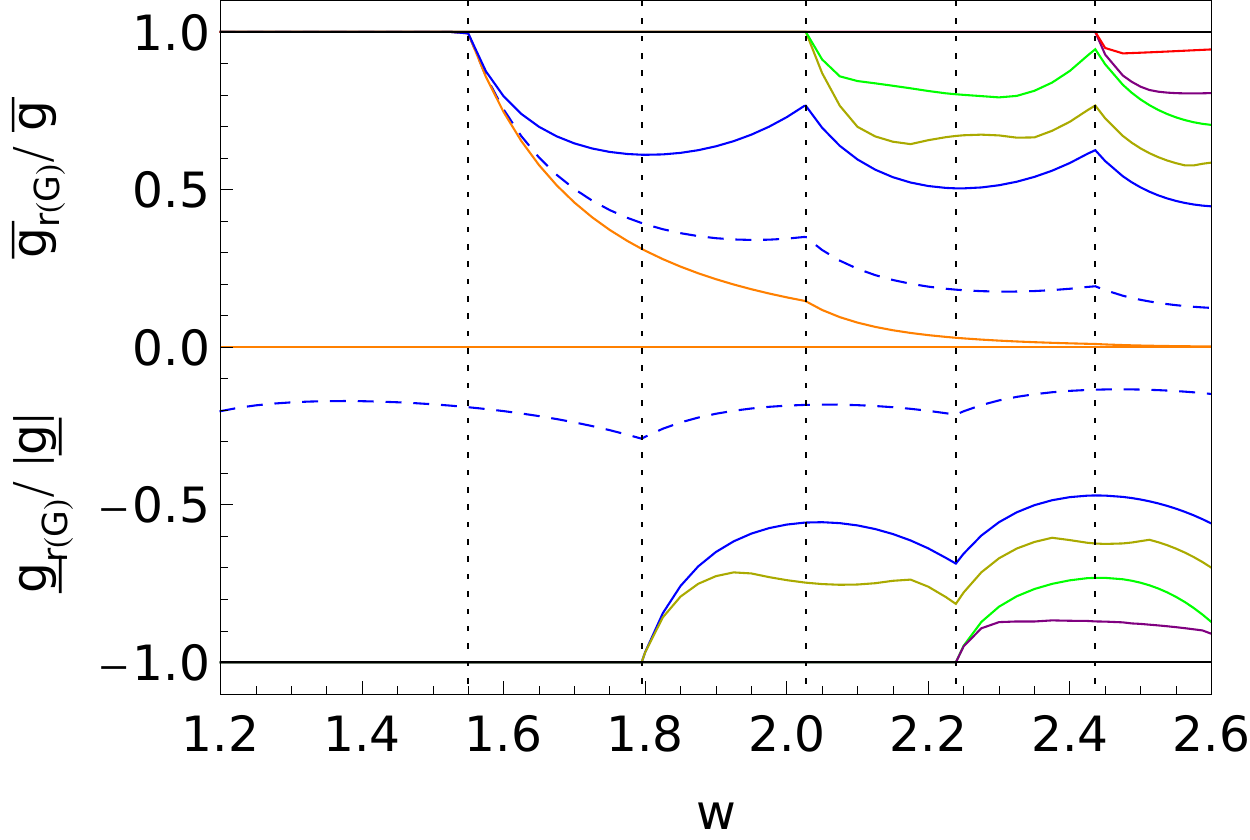}
		\caption{(Color online)
		The maximal (minimal) values $\overline{g}_r$ ($\underline{g}_r$) for $r=1,\dots,6$ as a function of the parameter $w$.
		Solid lines with positive (negative) function values from bottom to top (from top to bottom): $r=1$ (orange), $r=2$ (blue), $r=3$ (dark yellow), $r=4$ (green), $r=5$ (purple), and $r=6$ (red).
		We normalized the bounds to the overall supremum $\overline{g}$ (infimum $\underline{g}$).
 		The maximal (minimal) values $\overline{g}_{\mathrm{G}}$ ($\underline{g}_{\mathrm{G}}$) attained by any mixture of Gaussian states for a given $w$ are illustrated by the dashed blue lines, normalized in the same way.
 		The vertical dotted lines mark the critical values of $w$, where a next higher degree of nonclassicality $\kappa$ can be certified.
 	}\label{fig:levels}
	\end{figure}

	The result is shown in Fig. \ref{fig:levels} and gives a rich amount of insight.
	For better visualization of the relative effects, $\overline{g}_r(w)$ is normalized by the overall supremum, and $\underline{g}_r(w)$ by the modulus of the overall infimum of the nonclassicality quasiprobability. 
	One observes a splitting of the supremum (infimum) into three separated levels (trifurcation) whenever the increasing parameter $w$ traverses critical values, which turn out to be identical to the values where the overall supremum (infimum) is attained by another Fock state (see Table \ref{tab:wc}).
	This systematics most likely persists also for $w>2.6$, which allows us to arbitrarily increase the resolvable nonclassicality degree.
	The parameter $w$ should be chosen as large as necessary but as small as possible, as experimental data sampling noise increases with increasing $w$.
	
	Interestingly, in the case of $r=2$, the bounds $\overline{g}_r$ and $\underline{g}_r$ are attained by even and odd coherent states, $|\gamma_\pm\rangle=\mathcal{N}_\pm\left(|\gamma\rangle\pm|-\gamma\rangle\right)$, respectively, which are identified to be maximally nonclassical in the sense of (distance-based) topological nonclassicality measures. 
	The upper and lower bounds, $\overline{g}_{\mathrm{G}}$ and $\underline{g}_{\mathrm{G}}$, for the set $\mathcal{G}$ of mixtures of Gaussian states are also contained in Fig. \ref{fig:levels} and they are well separated from the levels $\overline{g}_r$ ($\underline{g}_r$); for further details on the optimal states see Appendix \ref{ch:optimalstates}.	
	Since $\mathcal{G}$ contains also the nonclassical squeezed states, it holds $\underline{g}_{\mathrm{G}}<\underline{g}_1=0$, and $\overline{g}_{\mathrm{G}}$ also exceeds the classical level $\overline{g}_1$.
	Note that, even for arbitrarily weak squeezing, these states are quantum superpositions of an infinite number of coherent states, referring to them as maximally nonclassical from the perspective of algebraic nonclassicality measures.
	Our results clearly show, that the nonclassicality quasiprobabilities uncover both the degree of nonclassicality and quantum non-Gaussianity without additional means.
		
\section{Photon-added squeezed states}
\label{ch:examples}

	\begin{figure}[b]
	\centering
		\begin{minipage}[t]{1.0\linewidth}
			(a)\\
			\includegraphics[width=0.95\linewidth]{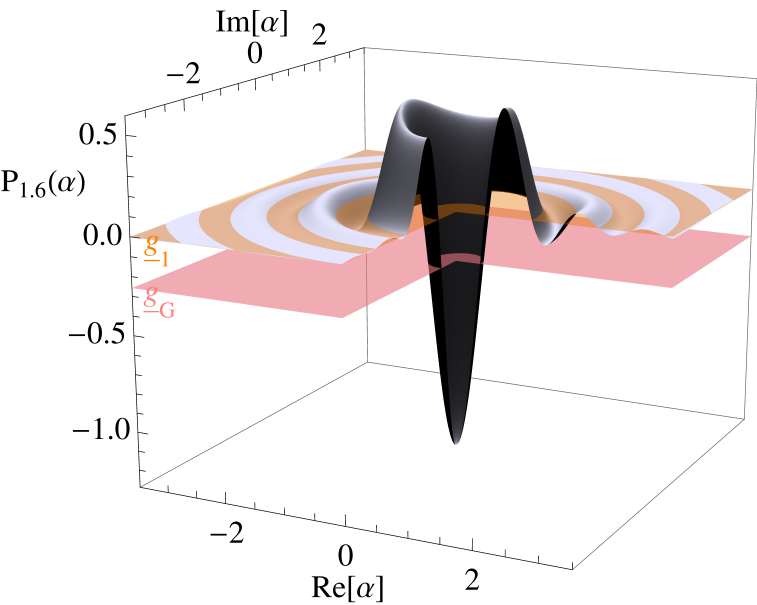}
			\label{fig:Reg1}
			\\\vspace*{3ex}(b)\\
			\includegraphics[width=0.95\linewidth]{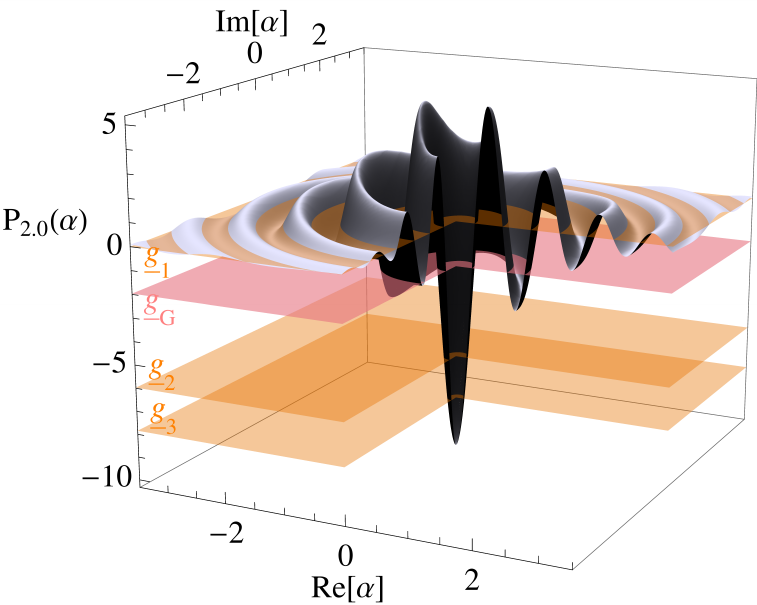}
			\label{fig:Reg3}
		\end{minipage}
		\caption{
			(Color online)
			Nonclassicality quasiprobability $P_w(\alpha)$ of an $m$-photon-added squeezed vacuum state for odd $m$ and squeezing parameter $\xi=0.1$:
			(a) $m=1$, $w=1.6$; (b) $m=3$, $w=2.0$.
			The lower bounds $\underline{g}_r$ for states with a nonclassicality degree of at most $r-1$ are given by the transparent orange planes.
			The lower bound for convex combinations of Gaussian states, $\underline{g}_{\mathrm{G}}$, is indicated by the transparent pink plane.
			For better visualization, we cut out the front-right quadrant.
		}
		\label{fig:examples}
	\end{figure}		

	The structure of the operator $\hat W_{w,\alpha}$ in Eq. \eqref{eq:Wwalpha} allows us to certify especially Fock-like states---states having a high overlap with a Fock state---to exceed the nonclassicality degree bounds $\overline{g}_r$ and $\underline{g}_r$.
	As an example, we consider the non-Gaussian states obtained by multiphoton additions to an initial Gaussian squeezed state, i.e.,
	\begin{align}\label{eq:added}
		|m,\xi\rangle=\mathcal{N}_{m,\xi}\,\hat a^{\dagger m}\hat S(\xi)|0\rangle,
	\end{align}
	with the unitary squeezing operator $\hat S(\xi)=\exp\left[\frac1{2}\left(\xi^\ast\hat a^2-\xi\hat a^{\dagger 2}\right)\right]$.
	The squeezing parameter is chosen to be $\xi=0.1$ and $m$ is the number of added photons.
	The prefactor $\mathcal{N}_{m,\xi}$ ensures the correct normalization.

	\begin{figure}[h]
	\centering
		\begin{minipage}[h]{1.0\linewidth}
			(a)\\
			\includegraphics[width=0.95\linewidth]{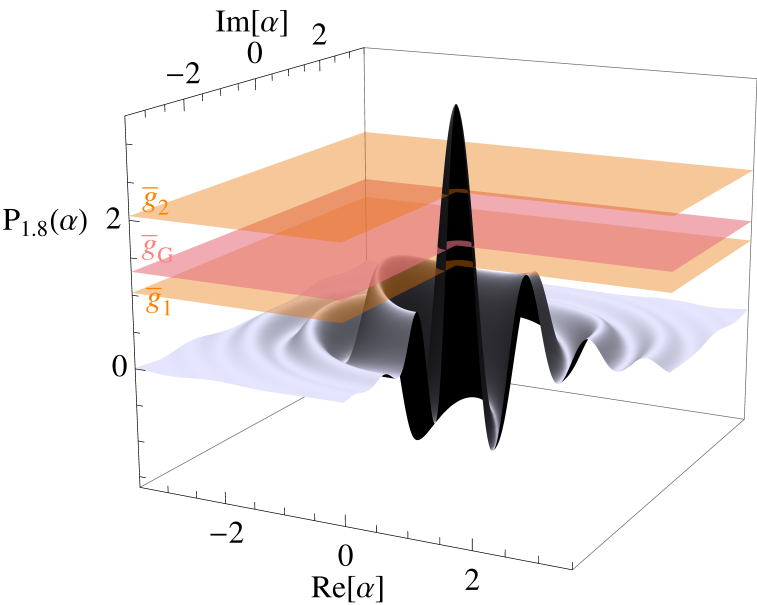}
			\label{fig:Reg2}
			\\\vspace*{3ex}	(b)\\
			\includegraphics[width=0.95\linewidth]{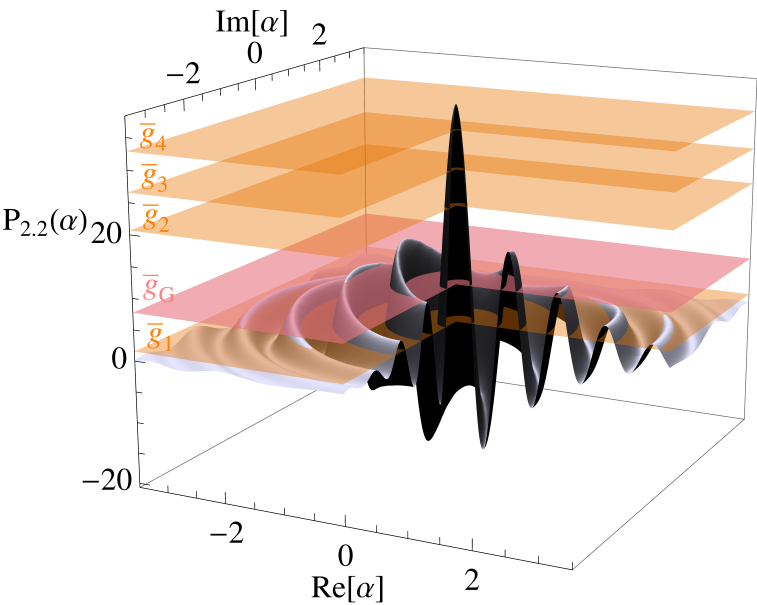}
			\label{fig:Reg4}
		\end{minipage}
		\caption{
			(Color online)
			Nonclassicality quasiprobability of an $m$-photon-added squeezed vacuum state for even $m$ and squeezing parameter $\xi=0.1$:
			(a) $m=2$, $w=1.8$; (b) $m=4$, $w=2.2$.
			The upper bounds $\overline{g}_r$ for states with a nonclassicality degree of at most $r-1$ are given by the transparent orange planes.
			The upper bound for convex combinations of Gaussian states, $\overline{g}_{\mathrm{G}}$, is indicated by the transparent pink plane.
			For better visualization, we cut out the front-right quadrant.
		}
		\label{fig:examples2}
	\end{figure}	
	
	Figures \ref{fig:examples}(a) and \ref{fig:examples}(b) show the nonclassicality quasiprobability of a single- and a three-photon-added squeezed vacuum state, respectively.
	For the single-photon addition, the phase-space function falls below $\underline{g}_1=0$ but not below $\underline{g}_2$.
	Thus, nonclassicality is certified ($\kappa>0$), but no information on the nonclassicality degree is extracted.   
	One clearly observes that through the addition of three photons the quasiprobability falls below the lower bound $\underline{g}_3$.
	Accordingly, this state has at least the nonclassicality degree $\kappa=3$; i.e., it is composed of a quantum superposition of at least four coherent states.
	Both states are certified to be quantum non-Gaussian ones, since $P_w(\alpha)$ penetrates $\underline{g}_{\mathrm{G}}$.
	
	In addition, we present in Figs.~\ref{fig:examples2}(a) and \ref{fig:examples2}(b) the results for photon-added squeezed vacuum states with two and four added photons and the same squeezing parameter as in Fig. \ref{fig:examples}.
	Since $P_w(\alpha)$ in Fig. \ref{fig:examples2}(a) exceeds the upper bound $\overline{g}_2$, the two-photon-added squeezed vacuum state is shown to have a nonclassicality degree of at least $2$.
	The nonclassicality quasiprobability in Fig. \ref{fig:examples2}(b) exceeds the upper bound $\overline{g}_4$ and, accordingly, the four-photon-added squeezed vacuum state has a nonclassicality degree of at least $4$.
	Both states strongly exceed the range allowed for convex combinations of Gaussian states; they are therefore shown to be quantum non-Gaussian.
			
\section{Conclusions}
\label{ch:conclusions}
	In conclusion, the nonclassicality quasiprobabilities were so far known to fully verify the nonclassicality of quantum states.
	Beyond this, we have shown that they additionally uncover important structural quantum characteristics of the state.
	Beneficially, both the degree of nonclassicality and the quantum non-Gaussianity are accessible via this single quantity.
	The nonclassicality degree, detected by our method, is useful for applications in quantum technologies, as the nonclassicality degree of a single-mode beam forwarded to the input of an $N$ splitter coincides with the amount of multipartite entanglement produced in the outputs.
	Most importantly, the structure of nonclassicality quasiprobabilities uncovers both algebraic and topological aspects of nonclassicality.\\
		
\begin{acknowledgements}
	The authors are grateful to S. Ryl and J. Sperling for enlightening discussions.
	This work has been supported by the European Commission through the project QCUMbER (Quantum Controlled Ultrafast Multimode Entanglement and Measurement), Grant No. 665148. 
\end{acknowledgements} 

\appendix

\section*{Appendix}

\section{Overall boundaries of nonclassicality quasiprobabilities}\label{ch:totalmaxmin}
	In this section the overall maximization and minimization of the nonclassicality quasiprobability $P_w(\alpha)$ with respect to all quantum states is studied.
	For reasons described in the main text, the overall supremum $\overline{g}(w)$ and infimum $\underline{g}(w)$ of the nonclassicality quasiprobability $P_w(\alpha)$ for a fixed value of $w$ is attained for Fock states $|\overline{n}(w)\rangle$ and $|\underline{n}(w)\rangle$, respectively, with
	\begin{equation}
	\begin{aligned}\label{eq:abs}
		\overline{n}(w)&=\arg \max_n\left[c_{w,n}\right],\notag\\
		\underline{n}(w)&=\arg \min_n\left[c_{w,n}\right],\notag\\
		\overline{g}(w)&=\max_n\left[c_{w,n}\right],\notag\\
		\underline{g}(w)&=\min_n\left[c_{w,n}\right],\notag
	\end{aligned}
	\end{equation}
	where the coefficients $c_{w,n}$ are defined in Eq. \eqref{eq:coeff} of the main text.
	
	\begin{figure}[h]
		\includegraphics[clip,scale=0.65]{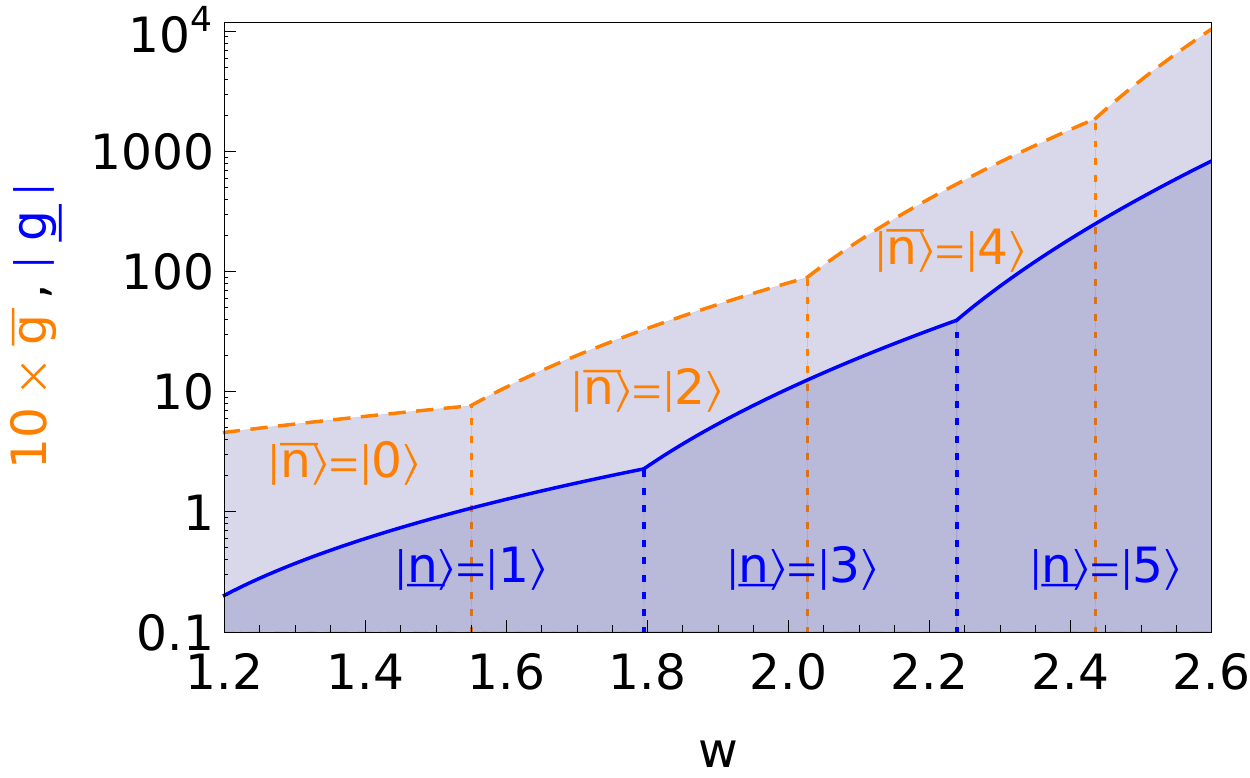}
		\caption{(Color online)
		The supremum $\overline g$ (dashed orange line) and the modulus of the infimum $\underline{g}$ (solid blue line) of $P_w$, which are attained by Fock states $|\overline{n}\rangle$ and $|\underline{n}\rangle$ as a function of the parameter $w$.
		Note that the supremum is depicted with a factor of $10$ for better visual separation of both curves. 
	}\label{fig:wmaxmin}
	\end{figure}
	
	These quantities are illustrated in Fig. \ref{fig:wmaxmin} on a logarithmic scale.
	Interestingly, one observes a well-ordered structure.
	The supremum is attained for Fock states with even number, while the infimum is attained for odd ones.
	The change of $\overline{n}$ ($\underline{n}$) at the critical values accompanies a discontinuity of the derivative of $\overline{g}(w)$ ($\underline{g}(w)$) with respect to $w$.
	
\section{Optimal states}\label{ch:optimalstates}
\begin{figure}[t]
		\includegraphics[width=0.9\linewidth]{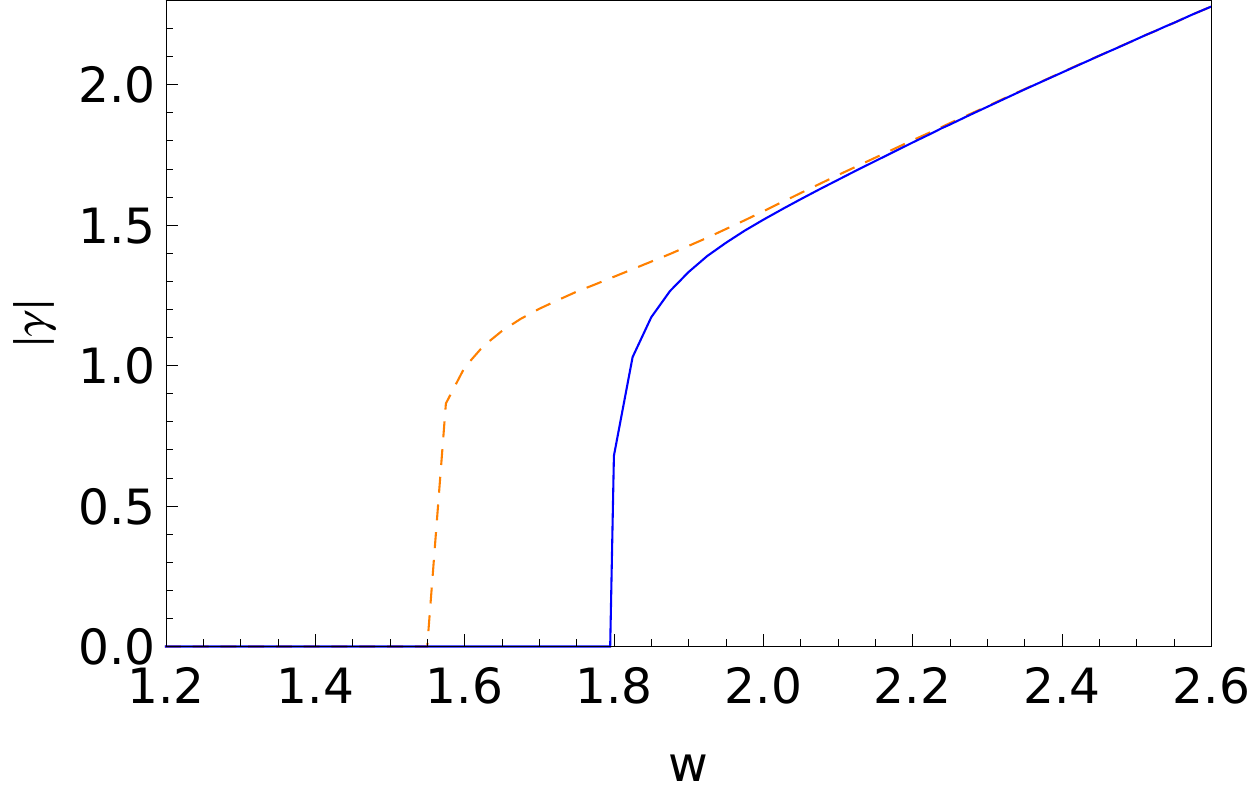}
		\caption{(Color online)
		Dashed orange line (solid blue line): Amplitude $|\gamma|$ of the even (odd) coherent state, which maximizes (minimizes) the nonclassicality quasiprobability $P_w(\alpha)$ with respect to the set $\mathcal{M}_2$ of states with a nonclassicality degree of $\kappa=1$ as a function of the parameter $w$.}
		\label{fig:amplitude}
	\end{figure}

	It turns out that even (odd) coherent states,
	\begin{align}\label{eq:cat}
		|\gamma_\pm\rangle=\dfrac1{\sqrt{2\left(1\pm e^{-2|\gamma|^2}\right)}}\left(|\gamma\rangle\pm|-\gamma\rangle\right),
	\end{align}
	where the ``$+$'' corresponds to even ones and the ``$-$'' to odd ones, maximize (minimize) the nonclassicality quasiprobability $P_w(\alpha)$ with respect to the set $\mathcal{M}_2$ of states with nonclassicality degree $\kappa=1$. 
	The absolute value of the amplitude $\gamma$, for which the maximum and minimum are obtained, is shown in Fig. \ref{fig:amplitude} as a function of $w$.
	The two curves  approach each other for increasing parameter $w$.
	Accordingly, for large $w$ the maximum and minimum are attained for even and odd coherent states with nearly the same amplitude $\gamma$.
	This optimal amplitude increases with increasing $w$.
	For $w<1.55$ the optimal amplitude of the even coherent state, corresponding to the maximum of $P_w(\alpha)$, is $\gamma=0$ in Eq. \eqref{eq:cat}, which is the vacuum state.
	For $w<1.795$ the amplitude of the odd coherent state providing the minimum is $\gamma=0$, which according to Eq. \eqref{eq:cat} coincides with the single-photon state.

	\begin{figure}[t]
		\includegraphics[width=0.9\linewidth]{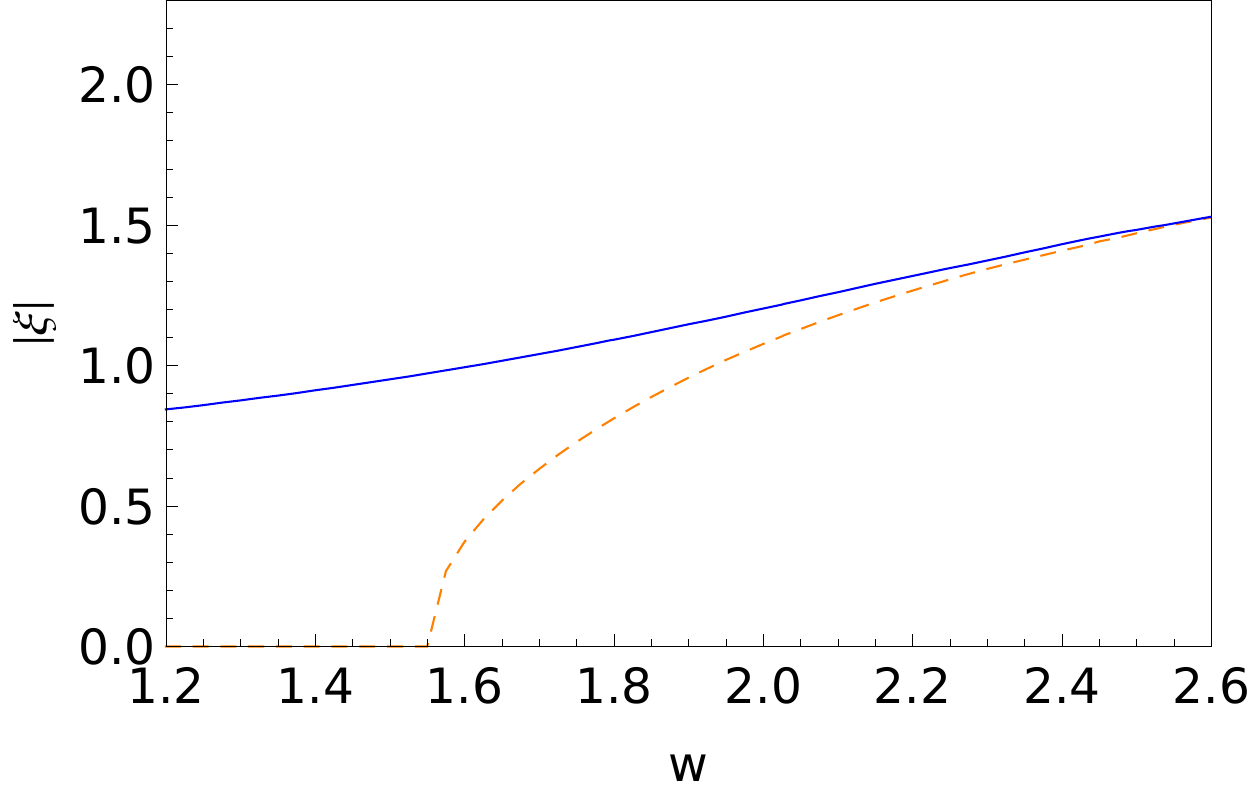}
		\caption{(Color online)
		Dashed orange line (solid blue line): Absolute value of the squeezing parameter $\xi$ of the squeezed vacuum state, which maximizes (minimizes) the nonclassicality quasiprobability $P_w(\alpha)$ with respect to the set $\mathcal{G}$ of mixtures of Gaussian states as a function of the parameter $w$.}
		\label{fig:squeezing}
	\end{figure}	
	
	The nonclassicality quasiprobability is optimized with respect to the set $\mathcal{G}$ of mixtures of Gaussian states for squeezed vacuum states [Eq. \eqref{eq:added} with $m=0$ in the main text].
	The absolute value of the squeezing parameter $\xi$, which yields the maximum and minimum of $P_w(\alpha)$, is illustrated in Fig. \ref{fig:squeezing} as a function of $w$.
	In the analyzed range of $w$, the minimum of the nonclassicality quasiprobability is attained for larger $|\xi|$ than the maximum.
	Furthermore, for $w<1.55$ one obtains the maximum for the vacuum state, which corresponds to $\xi=0$.
	We also observe that the maximum and minimum are obtained for the same increasing squeezing parameter in the limit of a large parameter $w$.

\end{document}